# Enhancing Lung Disease Diagnosis via Semi-Supervised Machine Learning

**Xiaoran Xu**[a] , **In-Ho Ra**[b] and * **Ravi Sankar**[c]

[a] iCONS Lab, Department of Electrical Engineering, University of South Florida, Tampa, FL, USA, xiaoranxu@usf.edu

[b] School of Software, Kunsan National University, Gunsan, South Korea, ihra@kunsan.ac.kr

[c] iCONS Lab, Department of Electrical Engineering, University of South Florida, Tampa, FL, USA, sankar@usf.edu

## Abstract

Lung diseases, including lung cancer and COPD, are significant health concerns globally. Traditional diagnostic methods can be costly, time-consuming, and invasive. This study investigates the use of semi-supervised learning methods for lung sound signal detection using a model combination of MFCC+CNN. By introducing semi-supervised learning modules such as Mix-Match, Co-Refinement, and Co-Refurbishing, we aim to enhance the detection performance while reducing dependence on manual annotations. With the add-on semi-supervised modules, the accuracy rate of the MFCC+CNN model is 92.9%, an increase of 3.8% to the baseline model. The research contributes to the field of lung disease sound detection by addressing challenges such as individual differences, feature complexity, and insufficient labeled data.

## 1 Introduction

Lung diseases are a group of serious ailments, including tuberculosis, lung cancer, asthma, and chronic obstructive pulmonary disease (COPD). Among these, lung cancer is one of the deadliest cancers worldwide, while COPD is the fourth leading cause of death globally. According to the World Health Organization [15], over 10 million people die from various lung diseases each year. Early detection and accurate diagnosis are critical for treating lung diseases, as early identification and treatment can prevent or delay disease progression and improve cure rates.

However, traditional lung disease diagnostic methods, such as CT scans [1], X-rays [6], and pulmonary function tests, can be costly and time-consuming. Moreover, these methods may cause radiation damage to patients. With the advancement of technology, an increasing number of researchers are exploring the use of lung sound signals for detecting lung diseases. Compared to traditional diagnostic methods, lung sound signal detection has the advantages of low cost, easy operation, and non-invasiveness. Therefore, it has become a popular non-invasive diagnostic method.

Nonetheless, the characteristics of lung sound signals are complex, challenging to extract accurately, and exhibit significant individual differences. These issues make the analysis and recognition of lung sound signals a challenging problem. Furthermore, traditional classification methods often require a large amount of manually annotated data, which is difficult to obtain in practice. As a result, semi-supervised learning has emerged as an effective solution to this problem.

In this context, the present study aims to investigate the use of semi-supervised learning methods to utilize unlabeled data, employing a model combination based on Mel-frequency cepstral coefficients (MFCC) [3] and convolutional neural networks (CNN) [5]. We will introduce semi-supervised learning modules, such as Mix-Match [2, 4], Co-Refinement [10], and Co-Refurbishing [10, 12], to enhance the detection performance of lung sound signals. By training deep learning models using labeled and unlabeled data, we can reduce dependence on manual annotations while maintaining classification performance and improving model generalization.

The main contribution of this research is to explore semi-supervised learning methods for utilizing unlabeled data to enhance lung sound signal detection performance and address challenges in the analysis and recognition of lung sound signals, such as individual differences, feature complexity, insufficient labeled data, and variability of sound patterns. By using deep learning models and semi-supervised learning modules, this study can reduce dependence on manual annotations, improve model generalization, and contribute to the research field of lung disease sound detection.

The remainder of this paper is organized as follows: Section 2 explains the related work; Section 3 presents the main research methods; Section 4 provides a detailed description of the experiments; Section 5 offers an analysis and evaluation of the experimental results; and the final section concludes and offers prospects for future research directions.

## 2 Related Work

In the field of audio, particularly in machine learning for lung disease sounds, there is substantial potential for further research in semi-supervised learning. Below are some related works [13].

Gupta et al. [4] proposed a framework that improves clustering performance iteratively by using ensemble methods to find high-quality pseudo-labels and training semi-supervised models. This framework surpassed existing techniques by 8-12% on CIFAR-10 and 20news datasets. Li et al. [9] introduced Divide-Mix, which uses mixture models to divide training data into clean and noisy samples and trains deep networks in a semi-supervised manner. They improved the Mix-Match strategy by refining and guessing labels on both labeled and unlabeled samples.

Berthelot et al. [2] developed Mix-Match, which guesses low-entropy labels for unlabeled examples and uses Mix-Up to blend labeled and unlabeled data. They reduced the error rate significantly on CIFAR-10 and STL-10. Song et al. [12] proposed SELFIE, a robust training method that selectively renovates and utilizes unclean samples that can be corrected with high accuracy. This method prevents noise accumulation and fully utilizes training data.

Van Gansbeke et al. [14] advocated a two-step method that decouples feature learning and clustering. First, they obtained semantically meaningful features using self-supervised tasks. Then, they used these features in a learnable clustering method. This approach eliminates the dependence of clustering learning on low-level features. They also introduced a semi-supervised deep learning algorithm for the automatic classification of lung sounds.

Lang et al. [7] proposed Graph Semi-Supervised CNNs (GS-CNNs) that classify respiratory sounds using a small number of labeled samples and many unlabeled samples. They constructed a respiratory sound graph (Graph-RS) and developed GS-CNNs by adding the information extracted from Graph-RS to the original CNN's loss function. This enhanced classification accuracy. Lang et al. [8] introduced a graph-based semi-supervised OCSVM using limited labeled samples to effectively detect abnormal lung sounds, enhancing recognition and generalization performance.

These studies demonstrate the potential of semi-supervised learning in lung disease sound analysis. They provide a basis for further exploration and development of novel methods and algorithms that can improve lung sound classification and detection. By leveraging semi-supervised learning techniques, researchers can overcome the limitations of supervised methods, enabling more efficient and accurate analysis of larger datasets with limited labeled samples, significantly contributing to machine learning in the field of audio, particularly for lung disease diagnosis and monitoring.

## 3 Approach

Through the research of related work, we choose to use the combination of MFCC+CNN to build the basic model [3, 5]. Incorporating three different modules on top of the baseline model for each training epoch, we first process the data using the Mix-Match class [2]. The Mix-Match function takes labeled data and unlabeled data as inputs and returns mixed samples and pseudo-labels after data augmentation and label guessing. Next, we train the model using these mixed samples and pseudo-labels. Additionally, we employ the Co-Refinement class to train the model, taking both the labeled and unlabeled data and training the model using the predicted results of the unlabeled data along with the labeled data. Finally, we use the Co-Refurbishing class to train the model, taking both the labeled and unlabeled data, and mix the predicted results of the unlabeled data with a subset of the labeled data to further improve the model's performance. The following is the basic model and the specific description of the modules used.

### 3.1 Baseline Model - MFCC+CNN

The MFCC plays a vital role in speech recognition applications and serves as a key feature to model the data and its spectrum [3]. It effectively captures the spectral characteristics of respiratory sounds, enabling accurate analysis and diagnosis. Originally developed for speech recognition, MFCC has been extended to lung sounds due to their similar time-varying nature and frequency-based features. The MFCC extraction process as shown in Figure 1 involves several steps: pre-emphasis, which enhances high-frequency parts for a more uniform frequency range; frame division, where the signal is split into short time frames assuming constant frequency within each frame; application of a window function, such as the Hamming window, to reduce spectral leakage; Fourier transformation to represent different voice characteristics through energy distribution in the frequency domain; triangular bandpass filtering for smoothing the frequency spectrum, noise elimination, formant highlighting, and computational reduction; logarithmic energy output calculation for each filter bank; and Discrete Cosine Transform (DCT) for decorrelation and compression of filter bank coefficients.

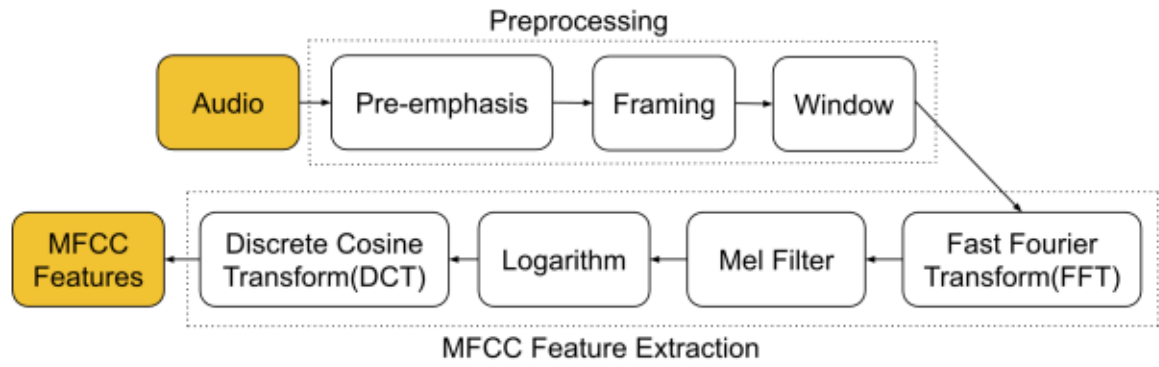

Figure 1: MFCC feature extraction processing.

Next is the structure of the CNN model [5]. In a sequential model composed of four Conv2D convolution layers, as shown in Figure 2, each layer utilizes a filter size of 2x2. The input shape for the first layer is (40, 862, 1), where 40 represents the number of Mel-frequency cepstral coefficients (MFCCs) and 862 corresponds to the number of frames. A modest dropout value of 20% is employed for the convolutional layers to minimize overfitting.

Each convolutional layer is paired with a MaxPooling2D pooling layer, with the final convolutional layer connected to a GlobalAveragePooling2D layer. The window size for the max pooling layers in the convolutional block is 2x2, and the stride is set to 2. The model's output is generated by a dense layer with 6 nodes (number of labels), which correspond to 6 distinct classes.

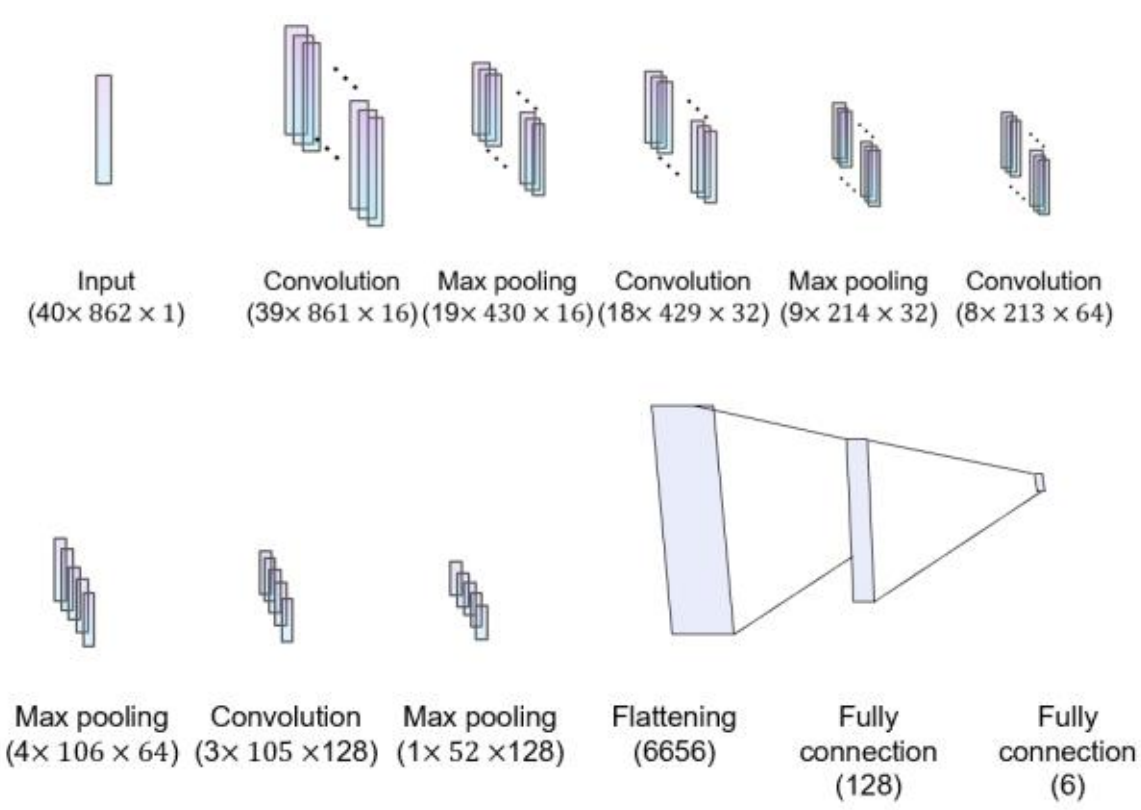

Figure 2: The structure of CNN.

### 3.2 Semi-supervised Module

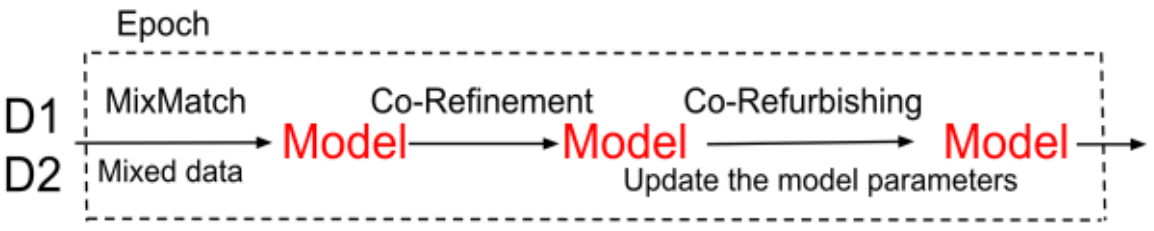

Figure 3: The module starts each epoch with the Mix-Match method, processing labeled (D1) and unlabeled (D2) data to obtain mixed data and labels for model training. Next, Co-Refinement uses predictions on unlabeled data as target labels, combined with labeled data for training. Finally, Co-Refurbishing updates model parameters by weighting prediction results of unlabeled data and true labels.

A. Mix-Match

This module implements Mix-Match, a semi-supervised learning method that leverages unlabeled data to enhance deep learning model performance [2,10]. Mix-Match performs data augmentation and label guessing on both labeled and unlabeled data. Key components include Mix-Up data augmentation, which generates new mixed samples through linear interpolation, and the Sharpen method, which adjusts predicted probability distributions to reduce uncertainty. The Mix-Match method is the main implementation, creating new training samples and labels from unlabeled data, enabling better model performance with limited labeled data.

B. Co-Refinement

The Co-Refinement module implements a semi-supervised learning approach using both labeled and unlabeled data to improve model performance. The train step method accepts labeled and unlabeled data, makes predictions on the unlabeled data, and merges them with labeled data for training. This approach allows the model to utilize both data types, improving performance with limited labeled data. Co-Refinement's primary advantage is its simplicity and ease of implementation.

C. Co-Refurbishing

The Co-Refurbishing module implements a semi-supervised learning approach that enhances deep learning model performance by leveraging unlabeled data [10,12]. This is achieved by making predictions on unlabeled data using the model, mixing these predictions with a subset of labeled data, and training the model with this combined data. Co-Refurbishing enables the model to utilize both labeled and unlabeled data simultaneously during training, improving performance with limited labeled data available. This method combines the advantages of supervised and self-supervised learning for better generalization when data is scarce.

# 4 Experiments

In this experiment, MFCC is used to extract spectral features from the audio data obtained from the ICHBI dataset [11]. These features are then fed into a CNN model, forming the base model. To improve the model's performance, three semi-supervised modules are incorporated, as described in the methodology section. For each epoch, the Mix-Match function is called, using labeled and unlabeled data to create a new combined dataset. This combined dataset is then used to train the model for one epoch. After training, two additional functions, Co-Refinement [10] and Co-Refurbishing [12], are called to perform extra training and optimization steps. Once the loop is completed, the model is trained again, this time using the full labeled dataset and the validation dataset.

Therefore, the training process initially trains the model using a combination of labeled and unlabeled data, followed by additional training or optimization steps before evaluating the model's accuracy and saving the model. Finally, the model is trained again using the complete labeled dataset and employs callback functions to control the training process.

The ICHBI dataset [11] contains sound recordings of respiratory conditions such as pneumonia, COPD, and asthma, gathered using advanced stethoscopes and electret amplifiers. The recordings were made in Aveiro and Porto, Portugal, as well as at the University of Southampton in England, from both clinical and non-clinical settings. The dataset includes 920 recordings from 126 participants and two sets of explanations. Recordings were taken from the windpipe and six chest areas, and participants had various respiratory conditions, including upper respiratory tract infections, pneumonia, COPD, asthma, bronchiolitis, bronchiectasis, and cystic fibrosis.

# 5 Evaluation and Analysis

During the evaluation of the complete model, we trained and validated on the labeled dataset, while also utilizing unlabeled data for semi-supervised learning. In the training process, we needed to balance the weights between labeled and unlabeled data to fully leverage the information from both sources. When evaluating the performance of the complete model, we focused on the same evaluation metrics as the base model.

| | precision | recall | f1-score | support |
|---|---|---|---|---|
| Bronchiectasis | 0.75 | 1.00 | 0.86 | 3 |
| Bronchiolitis | 0.50 | 0.67 | 0.57 | 3 |
| COPD | 0.97 | 0.96 | 0.97 | 159 |
| Healthy | 0.00 | 0.00 | 0.00 | 7 |
| Pneumonia | 0.36 | 0.71 | 0.48 | 7 |
| URTI | 0.25 | 0.20 | 0.22 | 5 |
| accuracy | | | 0.89 | 184 |
| macro avg | 0.47 | 0.59 | 0.52 | 184 |
| weighted avg | 0.88 | 0.89 | 0.88 | 184 |

Figure 4: Classification report of baseline model.

| | precision | recall | f1-score | support |
|---|---|---|---|---|
| Bronchiectasis | 1.00 | 0.33 | 0.50 | 3 |
| Bronchiolitis | 0.00 | 0.00 | 0.00 | 3 |
| COPD | 0.97 | 1.00 | 0.98 | 159 |
| Healthy | 0.50 | 0.43 | 0.46 | 7 |
| Pneumonia | 0.78 | 1.00 | 0.88 | 7 |
| URTI | 0.25 | 0.20 | 0.22 | 5 |
| accuracy | | | 0.93 | 184 |
| macro avg | 0.58 | 0.49 | 0.51 | 184 |
| weighted avg | 0.91 | 0.93 | 0.92 | 184 |

Figure 5: Classification report of baseline model add-on semi-supervised module.

We compared the performance of the baseline model and the complete model to understand the improvement in model performance with the add-on semi-supervised learning modules. We hoped to observe a significant performance improvement, especially in smaller datasets. There was a significant improvement in the COPD, pneumonia, and healthy classifications, while the results were less satisfactory for bronchiectasis and bronchiolitis, which had only 16 and 13 samples in total in the dataset, with only 13 and 10 respectively used for training, which did not allow the model to perform to its full potential. The precision and recall of pneumonia improved from 0.36 and 0.71 to 0.78 and 1, respectively, while the precision and recall of COPD, which had the most samples in the dataset, were 0.97 and 1, respectively, indicating good performance when sufficient data was available. In Figures 4 and 5, the high concentration of items on the diagonal line in the confusion matrix also confirmed the classification effectiveness of the semi-supervised learning module for each class. Thus, this method is effective in reducing the demand for labeled data, fully utilizing unlabeled data, and improving the performance of lung sound detection for pulmonary diseases.

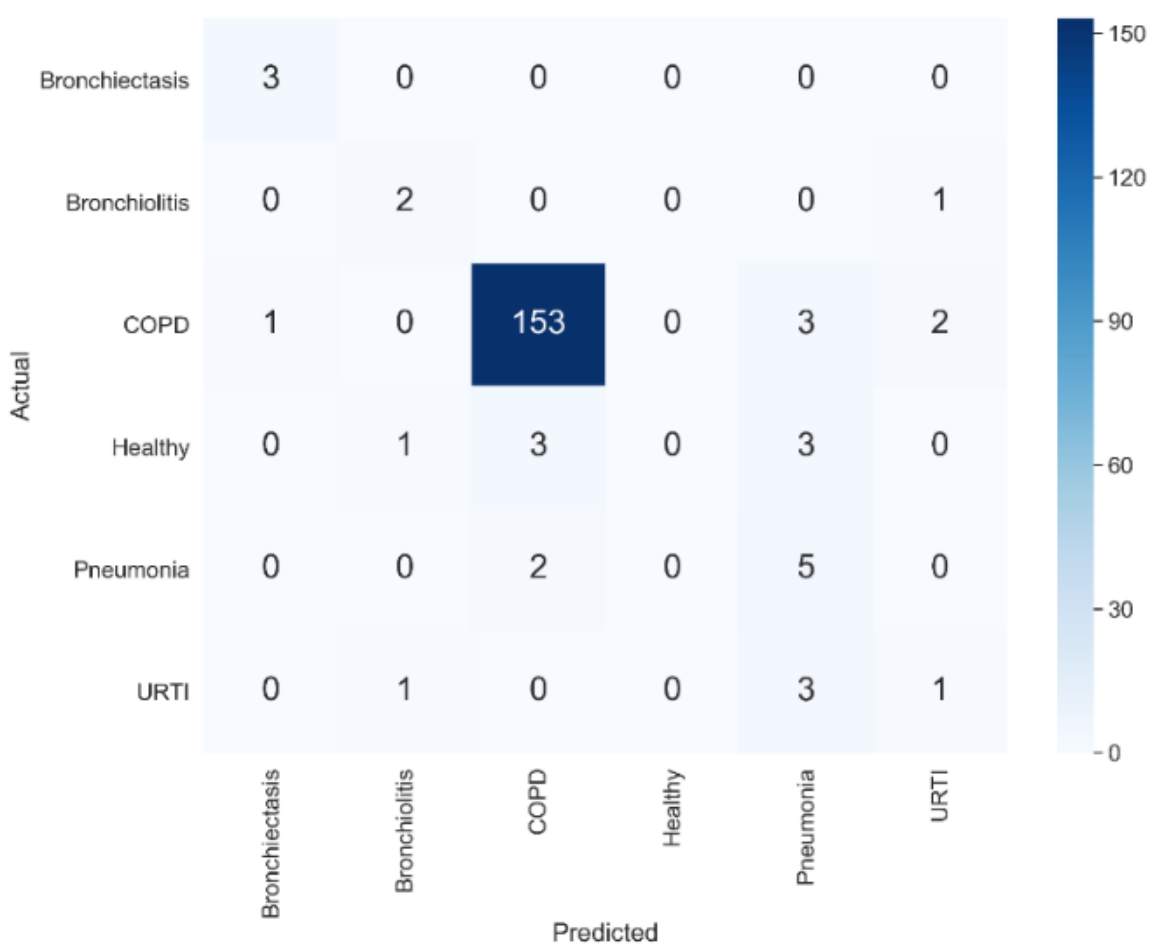

Figure 6: Confusion matrix of baseline model.

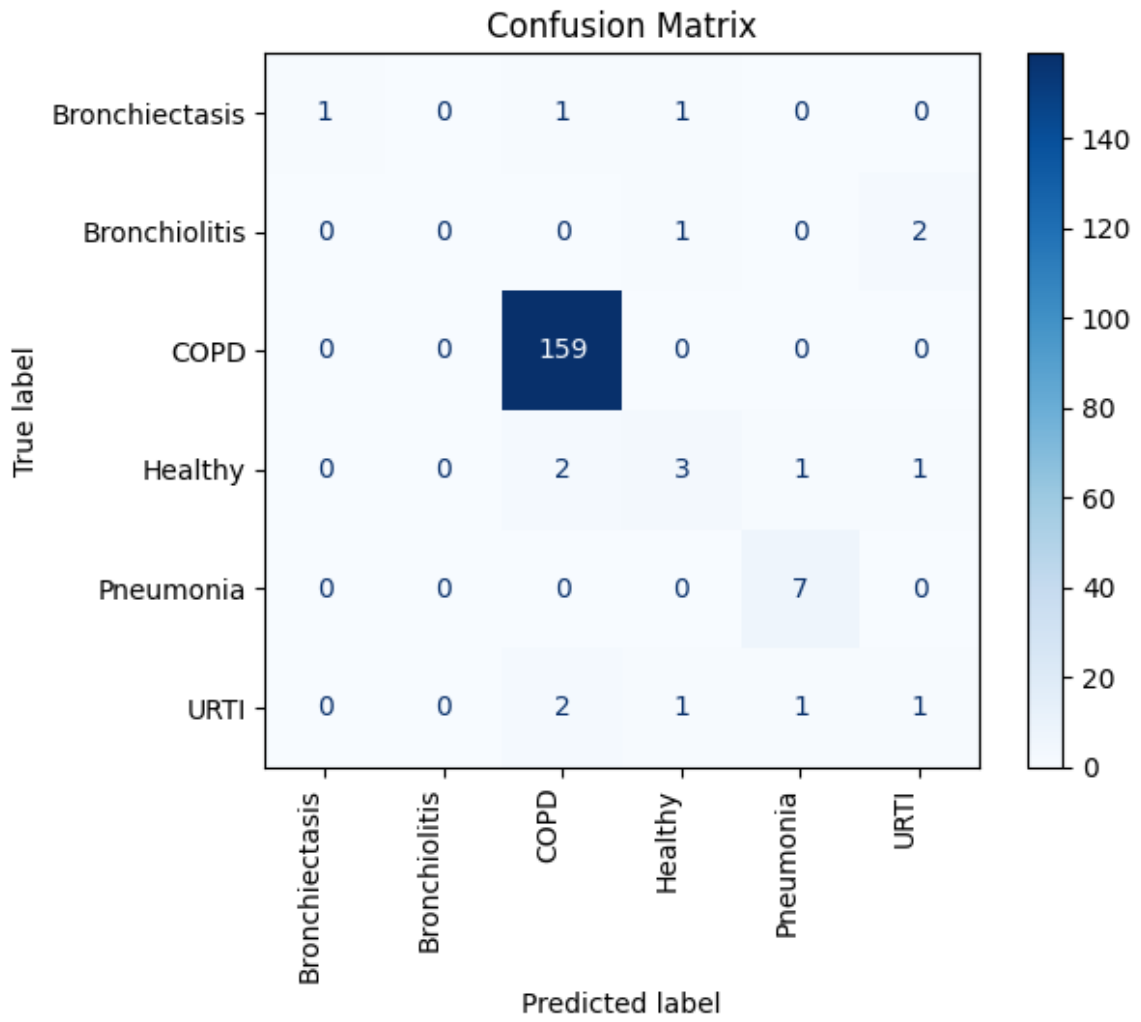

Figure 7: Confusion matrix of baseline model add-on semi-supervised module.

There are noticeable differences between the CoRefinement and CoRefurbishing methods when handling unlabeled data. The CoRefinement method directly employs the model's predictions on unlabeled data as target labels for training, whereas the CoRefurbishing method updates the model parameters by calculating a weighted average between the model's prediction results and the true labels of the labeled data. The CoRefinement method encourages the model to focus more on unlabeled data, while the CoRefurbishing method aids in rectifying the model's incorrect predictions on unlabeled data. In practical applications, the combination of these two methods results in superior performance. Upon removing the CoRefinement module, the accuracy is 89.7%. Similarly, when the CoRefurbishing module is removed, the accuracy is 90.7%. However, the final overall model achieves an accuracy of 92.9%. This demonstrates that the combination of both methods indeed leads to an improvement in performance, as per the results obtained from this experiment.

# 6 Conclusion

Leveraging semi-supervised learning methods, we developed an effective model for detecting lung disease sounds, enabling accurate and rapid preliminary diagnosis of various pulmonary sound diseases, thus helping doctors improve their efficiency. We utilized MFCC and CNN as the baseline model, with Mix-Match, Co-Refinement, and Co-Refurbishing serving as semi-supervised learning modules. Compared to the baseline model with the add-on semi-supervised modules, the accuracy was significantly improved from 89.1% to 92.9%.

Semi-supervised learning not only enhances the model's overall accuracy and generalization capabilities but also demonstrates substantial improvements in smaller datasets, such as the Pneumonia and Healthy classifications. This suggests that the use of semi-supervised learning reduces the need for labeled data, effectively utilizing unlabeled data and significantly lowering data requirements, thus allowing limited pulmonary disease data to have a greater impact.

This experiment is an initial attempt, but the preliminary results indicate promising outcome. In the future, we hope to compare a broader range of datasets and explore applications on other models to further verify the reliability and practicality of this approach.

# Acknowledgement

This work was supported by the National Research Foundation of Korea (NRF) grant funded by the Korea government (MSIT) (No. 2021R1A2C2014333)